# Análise Demográfica e Socioeconômica do Uso e do Acesso a Medicamentos Antidepressivos no Brasil

*Karinna Moura Boaviagem[1]*

*José Ricardo Bezerra Nogueira[2]*

(novembro de 2021)

**RESUMO**

Os transtornos depressivos, além de causar impactos negativos diretos à saúde, também são responsáveis por impor custos substanciais à sociedade. Em relação ao tratamento da depressão, os antidepressivos têm se mostrado eficazes e, para a Organização Mundial da Saúde, o acesso a psicotrópicos para pessoas com doenças mentais oferece uma chance de melhora da saúde e uma oportunidade de reengajamento na sociedade. O objetivo deste estudo é analisar o uso e o acesso a antidepressivos no Brasil, segundo as macrorregiões e às condições demográficas, sociais e econômicas da população, utilizando a Pesquisa Nacional de Acesso, Uso e Promoção do Uso Racional de Medicamentos (PNAUM 2013/2014). Os resultados mostram que há alta prevalência de uso de antidepressivos em indivíduos com depressão no Brasil. O perfil preponderante de uso desses medicamentos é: pessoas do sexo feminino, entre 20 e 59 anos, brancas, da região Sudeste, da classe econômica D/E, com nível de ensino médio, em situação conjugal, sem cobertura de plano de saúde, sem limitações derivadas da depressão, e quem se autoavaliou a saúde como regular.

**Palavras-chave**: Antidepressivos. Uso de medicamentos. Acesso a medicamentos. Fatores socioeconômicos.

**ABSTRACT**

Depressive disturbs, in addition to causing direct negative impacts on health, are also responsible for imposing substantial costs on society. In relation to the treatment of depression, antidepressants have proven effective, and, to the World Health Organization, access to psychotropic drugs for people with mental illnesses offers a chance of improved health and an opportunity for reengagement in society. The aim of this study is to analyze the use of and access to antidepressants in Brazil, according to macro-regions and to demographic, social and economic conditions of the population, using the National Survey on Access, Use and Promotion of Rational Use of Medicines (PNAUM 2013/2014). The results show that there is a high prevalence of antidepressant use in individuals with depression in Brazil. The main profile of use of these drugs is: female individuals, between 20 and 59 years old, white, from the Southeast region, of the economic class D/E, with a high schooling level, in a marital situation, without health insurance coverage, without limitations derived from depression, and who self-evaluated health as a regular.

**Keywords**: Antidepressants. Use of medications. Access to medicines. Socioeconomic factors.

---

[1] Secretaria Municipal de Saúde, Jaboatão dos Guararapes, Gerência de Assistência farmacêutica, e-mail: kboaviagem@gmail.com.

[2] Departamento de Economia, Universidade Federal de Pernambuco, e-mail: jrbnogueira@yahoo.com.br.

**1. Introdução**

Estudos de prevalência mostram que depressão é um transtorno frequente em vários países, com uma taxa de prevalência anual entre 3% e 11% (FLECK et al., 2009). No Brasil, depressão foi considerada uma das doenças crônicas mais frequentes na população, com base nos dados da Pesquisa Nacional de Saúde 2013, atingindo 7,6% da população (MALTA et al., 2015). No que diz respeito ao tratamento, "existe uma evidência contundente na literatura de que os medicamentos antidepressivos são eficazes no tratamento da depressão aguda de moderada a grave" (FLECK et al., 2009, p. 9).

Segundo a Organização Mundial de Saúde (OMS), "o acesso a medicamentos psicotrópicos para pessoas com doenças mentais oferece a chance de melhoria na saúde e a oportunidade de reengajamento na sociedade" (WORLD HEALTH ORGANIZATION, 2017). Entretanto, no Brasil, embora o acesso a medicamentos seja considerado alto, desigualdades socioeconômicas ainda são observadas no que diz respeito a esse acesso (BERTOLDI et al., 2009; HOGERZEIL e MIRZA, 2011).

No presente trabalho, utilizando microdados da PNAUM 2013/2014, investigou-se, para o Brasil, se há uma associação entre o uso e o acesso a medicamentos antidepressivos (AD) e as condições demográficas e socioeconômicas da população.

**2. Dados**

Este trabalho utilizou microdados da Pesquisa Nacional de Acesso, Utilização e Promoção do Uso Racional de Medicamentos (PNAUM), 2013/2014. A amostra da PNAUM consiste em 41.433 moradores em domicílios permanentes na zona urbana, em municípios das 26 unidades da Federação brasileira e no Distrito Federal (BRASIL, 2016).

Dentre os entrevistados foram selecionados aquelas pessoas que responderam positivo para a seguinte questão: "Algum médico já lhe disse que o(a) Sr(a) tem depressão?". A partir dessa amostra da população, foram analisados, como variáveis dependentes (a serem explicadas) o uso e o acesso a medicamentos antidepressivos.

O uso de medicamentos para tratamento de depressão foi identificado por meio das questões: "O(a) Sr(a) tem indicação médica para usar algum remédio para a depressão?" e "O(a) Sr(a) está usando algum desses remédios?". O acesso aos medicamentos foi identificado por meio das perguntas: "Entrevistado obtém algum dos remédios que utiliza no SUS?"; "Neste local o(a) Sr(a) consegue todos os remédios que precisa?"; "Entrevistado obtém algum dos remédios que utiliza em farmácia(s) privada(s)?" e "Neste local o(a) Sr(a) consegue todos os remédios que precisa?".

As variáveis independentes (explicativas) utilizadas foram as seguintes características demográficas e socioeconômicas: sexo, idade, região do Brasil, Critério de Classificação Econômica Brasil (CCEB), situação conjugal, raça, escolaridade e plano de saúde.

Para a variável idade, foram consideradas as seguintes faixas etárias: 15 a 19 anos; 20 a 59 anos e 60 anos ou mais. O CCEB se caracteriza como um escore baseado em um sistema de pontos atribuídos a algumas variáveis, como, por exemplo, bens que estão dentro do domicílio em funcionamento. Os cortes utilizados foram: classe A (45-100 pontos); classe B1 (38-44 pontos); classe B2 (29-37 pontos); classe C1 (23-28 pontos); classe C2 (17-22 pontos) e classe D/E (1-16 pontos). Para a situação conjugal, foi considerada a resposta "não" como o agrupamento entre as opções "não, mas já viveu antes" e "não viveu".

## 3. Método

A investigação realizada foi do tipo observacional, de corte transversal, descritivo, analítico, de base populacional e com abordagem quantitativa. Regressões logísticas (univariada e multivariada) foram realizadas para investigar os fatores associados ao uso e ao acesso de antidepressivos (AD).

Utilizou-se o método de regressão logística binária, com duas variáveis dependentes dicotômicas: uso de medicamentos e acesso a medicamentos. A medida de associação foi a razão de chances (*Odds Ratio* – OR), que avalia a relação entre a chance de um indivíduo exposto possuir a condição de interesse, comparada com a de um indivíduo não exposto. O OR foi expresso na forma de intervalo de confiança, calculado a partir de uma margem de erro pré-determinada (RUMEL, 1986). Foi escolhido o intervalo de confiança no nível de 95% (IC 95%).

O modelo univariado foi usado para determinar as variáveis mais significativamente associadas ao desfecho de interesse e, assim, selecionar as variáveis a serem incluídas na análise multivariada, onde as variáveis selecionadas são investigadas, simultaneamente, em relação às suas correlações com as variáveis dependentes (PAES, 2010).

Para uso no modelo multivariado, foram selecionadas aquelas variáveis independentes que apresentaram um p valor igual ou menor que 0,20 (p ≤ 0,20) em relação ao desfecho. Um nível de significância p < 0,05 foi o critério adotado para identificar as características independentemente associadas ao uso e ao acesso a medicamentos.

## 4. Resultados

Primeiramente, apresenta-se uma descrição geral da população analisada em termos de frequências absolutas e relativas das suas características demográficas e socioeconômicas. Posteriormente, são apresentados os resultados da análise de inferência da correlação entre as características demográficas e socioeconômicas e o acesso e uso de medicamentos antidepressivos.

### 4.1. Análise descritiva

A amostra da PNAUM selecionada foi composta por 32.652 indivíduos. Com a aplicação dos pesos amostrais, tem-se uma população geral estudada de 130.741.284 indivíduos. A Tabela 1 mostra as frequências absolutas e relativas da população, segundo as condições demográficas e socioeconômicas.

Tabela 1 – Frequências absolutas e relativas da população segundo as condições demográficas e socioeconômicas.

| Variável e Categorias | Frequências (%) |
|---|---|
| **Sexo** | |
| Masculino | 60.677.695 (46,4%) |
| Feminino | 70.063.590 (53,6%) |
| **Faixa Etária** | |
| 15 a 19 anos | 12.979.908 (9,9%) |
| 20 a 59 anos | 96.460.130 (73,8%) |
| 60 anos ou mais | 21.301.246 (16,3) |
| **Região do Brasil** | |
| Norte | 9.080.814 (6,9%) |
| Nordeste | 30.575.088 (23,4%) |
| Sudeste | 61.731.853 (47,2%) |
| Sul | 19.041.450 (14,6%) |
| Centro-Oeste | 10.312.079 (7,9%) |
| **CCEB**[a] | |
| Classe A | 0 (0%) |
| Classe B1 | 210.220 (0,2%) |
| Classe B2 | 6.343.339 (4,9%) |
| Classe C1 | 24.614.158 (19%) |
| Classe C2 | 46.370.042 (35,8%) |
| Classe D/E | 51.837.703 (40,1%) |
| **Situação conjugal** | |
| Sim | 70.336.466 (53,8%) |
| Não | 54.360.702 (41,6%) |
| NS/NR | 6.044.117 (4,6%) |
| **Raça** | |
| Branca | 56.535.495 (43,2%) |
| Negra | 11.718.586 (9%) |
| Amarela | 1.503.978 (1,1%) |
| Parda | 52.755.824 (40,4%) |
| Indígena | 461.106 (0,4%) |
| NS/NR | 7.766.295 (5,9%) |
| **Escolaridade** | |
| Curso primário | 23.176.772 (20,4%) |
| Admissão | 1.023.343 (1%) |
| Curso ginasial ou ginásio | 5.952.033 (5,2%) |
| 1º grau ou fundamental ou supletivo de primeiro grau ou EJA | 30.147.928 (26,6%) |
| 2º grau ou colégio técnico/normal/científico/ensino médio/supletivo de segundo grau/EJA | 36.881.625 (32,5%) |
| 3º grau ou curso superior | 12.602.343 (11,1%) |
| Pós-graduação | 2.111.826 (1,9%) |
| Nunca estudou | 645.456 (0,6%) |
| NS/NR | 832.921 (0,7%) |
| **Plano de saúde** | |
| Sim | 30.034.435 (23%) |
| Não | 100.590.974 (76,9) |
| Não sabe | 97.279 (0,1%) |

Fonte: Elaboração própria a partir dos microdados da PNAUM 2013/2014.

[a] Critério de Classificação Econômica Brasil (2013).

Legenda: NS/NR – não sabe/não respondeu

Na Tabela 2 encontram-se descritas as prevalências relacionadas à depressão autorreferida na PNAUM.

Tabela 2 – Prevalências autorreferidas de depressão, indicação médica de tratamento farmacológico e uso de medicamentos antidepressivos.

| N | Depressão autorreferida N (%) | Portadores de depressão com indicação de tratamento farmacológico N (%) | Portadores de depressão em uso de medicamentos antidepressivos N (%) |
|---|---|---|---|
| 130.741.284 | 7.051.362 (5,4%) | 5.697.662 (4,4%) | 5.115.580 (3,9%) |

Fonte: Elaboração própria a partir dos microdados da PNAUM 2013/2014.

A Tabela 3 descreve as características da população com depressão autorreferida em relação às variáveis demográficas e socioeconômicas.

Tabela 3 – Prevalências de portadores de depressão em uso de medicamentos AD segundo características demográficas e socioeconômicas. PNAUM 2013/2014.

| Variável e categorias | Portadores de depressão em uso de medicamentos antidepressivos N (%) |
|---|---|
| **Sexo** | |
| Masculino | 1.132.831 (19,9%) |
| Feminino | 4.564.831 (80,1%) |
| **Faixa Etária** | |
| 15 a 19 anos | 111.497 (2%) |
| 20 a 59 anos | 3.966.427 (69,6%) |
| 60 anos ou mais | 1.619.738 (28,4%) |
| **Região do Brasil** | |
| Norte | 122.769 (2,2%) |
| Nordeste | 1.036.654 (18,2%) |
| Sudeste | 2.826.899 (49,6%) |
| Sul | 1.328.134 (23,3%) |
| Centro-Oeste | 383.207 (6,7%) |
| **CCEB[a]** | |
| Classe A | 0 |
| Classe B1 | 4.728 (0,1%) |
| Classe B2 | 365.687 (6,5%) |
| Classe C1 | 1.060.785 (18,9%) |
| Classe C2 | 1.990.700 (35,4%) |
| Classe D/E | 2.199.325 (39,1%) |
| **Situação conjugal** | |
| Sim | 3.264.737 (57,3%) |
| Não | 2.401.429 (42,1%) |
| NS/NR | 31.496 (0,6%) |
| **Raça** | |
| Branca | 3.214.257 (56,4%) |
| Negra | 399.972 (7%) |
| Amarela | 55.776 (1%) |
| Parda | 1.908.478 (33,5%) |
| Indígena | 9.244 (0,2%) |
| NS/NR | 109.935 (1,9%) |
| **Escolaridade** | |
| Curso primário | 988.883 (20,13%) |
| Admissão | 31.272 (0,64%) |
| Curso ginasial ou ginásio | 254.215 (5,17%) |
| 1º grau ou fundamental ou supletivo de primeiro grau ou EJA | 1.311.565 (26,7%) |
| 2º grau ou colégio técnico/normal/científico/ensino médio/supletivo de segundo grau/EJA | 1.586.844 (32,3%) |
| 3º grau ou curso superior | 566.821 (11,54%) |
| Pós-graduação | 78.746 (1,6%) |
| Nunca estudou | 33.775 (0,7%) |
| NS/NR | 59.969 (1,22%) |
| **Plano de saúde** | |
| Sim | 1.719.324 (30,17%) |
| Não | 3.976.690 (69,8%) |
| Não sabe | 1.649 (0,03%) |

Fonte: Elaboração própria a partir dos microdados da PNAUM 2013/2014.

[a] Critério de Classificação Econômica Brasil (2013).

Legenda: AD – antidepressivos; NS/NR – não sabe/não respondeu

A Tabela 4 apresenta o perfil demográfico e socioeconômico relacionado ao acesso a medicamentos antidepressivos.

Tabela 4 – Acesso a AD por portadores de depressão em uso de medicamentos, segundo características demográficas e socioeconômicas da amostra. PNAUM 2013/2014.

| Variável | Acesso Farmácia Pública - SUS | Acesso Farmácia Privada |
|---|---|---|

|  | N (%) | N (%) |
|---|---|---|
| **Sexo** | | |
| Masculino | 664.898 (18,8%) | 724.291 (18%) |
| Feminino | 2.860.987 (81,2%) | 3.301.368 (82%) |
| **Faixa Etária** | | |
| 15 a 19 anos | 49.996 (1,42%) | 83.819 (2,08%) |
| 20 a 59 anos | 2.502.478 (70,97%) | 2.735.447 (67,95%) |
| 60 anos ou mais | 1.206.392 (27,61%) | 1.206.392 (29,97%) |
| **Região do Brasil** | | |
| Norte | 65.953 (1,87%) | 84.506 (2,1%) |
| Nordeste | 549.164 (15,57%) | 848.780 (21,08%) |
| Sudeste | 1.826.505 (51,81%) | 1.812.399 (45,02%) |
| Sul | 887.429 (25,17%) | 1.011.630 (25,13%) |
| Centro-Oeste | 196.835 (5,58%) | 268.343 (6,67%) |
| **CCEB[a]** | | |
| Classe A | 0 (0%) | 0 (0%) |
| Classe B1 | 2.101 (0,06%) | 4.728 (0,12%) |
| Classe B2 | 121.837 (3,48%) | 361.564 (9,11%) |
| Classe C1 | 574.383 (16,43%) | 851.723 (21,46%) |
| Classe C2 | 1.306.919 (37,37%) | 1.447.044 (36,46%) |
| Classe D/E | 1.491.539 (42,66%) | 1.303.516 (32,85%) |
| **Situação conjugal** | | |
| Sim | 2.005.234 (56,87%) | 2.408.878 (59,84%) |
| Não | 1.497.107 (42,46%) | 1.609.495 (39,98%) |
| NS/NR | 23.544 (0,67%) | 7.286 (0,18%) |
| **Raça** | | |
| Branca | 1.929.467 (54,73%) | 2.393.037 (59,44%) |
| Negra | 251.164 (7,12%) | 22.024 (5,56%) |
| Amarela | 38.168 (1,08%) | 36.527 (0,91%) |
| Parda | 1.237.499 (35,1%) | 1.287.643 (31,98%) |
| Indígena | 3.293 (0,09%) | 7.175 (0,18%) |
| NS/NR | 66.294 (1,88%) | 77.253 (1,92%) |
| **Escolaridade** | | |
| Curso primário | 652.460 (21%) | 648.354 (18,87%) |
| Admissão | 12.561 (0,40%) | 15.081 (0,44%) |
| Curso ginasial ou ginásio | 140.545 (4,5%) | 213.789 (6,22%) |
| 1º grau ou fundamental ou supletivo de primeiro grau ou EJA | 819.153 (26,3%) | 927.703 (27%) |
| 2º grau ou colégio técnico/normal/científico/ensino médio/supletivo de segundo grau/EJA | 1.025.176 (32,9%) | 1.071.831 (31,19%) |
| 3º grau ou curso superior | 383.956 (12,3%) | 437.341 (12,73%) |
| Pós-graduação | 39.747 (1,3%) | 69.619 (2,03%) |
| Nunca estudou | 18.428 (0,6%) | 14.806 (0,43%) |
| NS/NR | 20.777 (0,7%) | 37.922 (1,10%) |
| **Plano de saúde** | | |
| Sim | 751.788 (21,32%) | 1.464.058 (36,37%) |
| Não | 2.774.098 (78,68%) | 2.559.951 (63,59%) |
| Não sabe | 0 (0%) | 1.649 (0,04%) |

Fonte: Elaboração própria a partir dos microdados da PNAUM 2013/2014.

[a] Critério de Classificação Econômica Brasil (2013).

Legenda: AD – antidepressivos; NS/NR – não sabe/não respondeu

## 4.2. Análise de regressão logística

Abaixo apresenta-se, separadamente, os resultados das regressões logísticas realizadas em relação ao uso de medicamentos antidepressivos e ao acesso a esses medicamentos.

### 4.2.1 Análise relacionada ao uso de medicamentos AD

A Tabela 5 apresenta os resultados das análises univariada (*OR não ajustada*) e multivariada (*OR ajustada*) expressos em termos de razões de chances (OR) associadas ao uso de medicamentos em relação às características demográficas e socioeconômicas.[3]

---

[3] Ressalta-se que as categorias "classe A" e "classe B1" da variável *CCEB* foram excluídas, considerando que não houve casos registrados para a primeira categoria (0%) e a segunda representou apenas 0,1% da amostra. Dessa forma, para efeito de melhor comparabilidade, optou-se pela exclusão das referidas categorias da análise, tomando a "classe B2" como a referência. Da mesma forma, a categoria "não sabe" da variável *plano de saúde* também foi excluída, considerando que apresentou apenas 0,03% da amostra. Portanto, optou-se por excluí-la visando a comparação de interesse apenas entre aqueles indivíduos que possuíam e os que não possuíam convênio.

Tabela 5 – Análises univariada (*OR não ajustada*) e multivariada (*OR ajustada*) dos fatores demográficos e socioeconômicos associados ao uso de AD.

| Variável e Categorias | Razão de chances (OR) não ajustada | IC 95% | p | Razão de chances (OR) ajustada | IC 95% | p |
|---|---|---|---|---|---|---|
| **Sexo** | | | | | | |
| Masculino | 1,000 | | | 1,000 | | |
| Feminino | 1,691 | 1,681 - 1,701 | 0,00 | 1,702 | 1,708 - 1,732 | 0,00 |
| **Faixa Etária** | | | | | | |
| 15 a 19 anos | 1,000 | | | 1,000 | | |
| 20 a 59 anos | 3,259 | 3,216 - 3,302 | 0,00 | 6,829 | 6,717 - 6,942 | 0,00 |
| 60 anos ou mais | 8,128 | 8,010 - 8,247 | 0,00 | 13,235 | 13,005 - 13,470 | 0,00 |
| **Região do Brasil** | | | | | | |
| Norte | 1,000 | | | 1,000 | | |
| Nordeste | 1,444 | 1,423 - 1,465 | 0,00 | 1,023 | 1,005 - 1,040 | 0,011 |
| Sudeste | 3,104 | 3,061 - 3,147 | 0,00 | 1,773 | 1,744 - 1,803 | 0,00 |
| Sul | 4,267 | 4,203 - 4,331 | 0,00 | 2,382 | 2,339 - 2,426 | 0,00 |
| Centro-Oeste | 2,375 | 2,336 - 2,415 | 0,00 | 1,437 | 1,409 - 1,466 | 0,00 |
| **CCEB[a]** | | | | | | |
| Classe B2 | 1,000 | | | 1,000 | | |
| Classe C1 | 1,153 | 1,138 - 1,168 | 0,00 | 1,837 | 1,811 - 1,863 | 0,00 |
| Classe C2 | 1,247 | 1,232 - 1,261 | 0,00 | 1,941 | 1,915 - 1,967 | 0,00 |
| Classe D/E | 0,737 | 0,728 - 0,745 | 0,00 | 1,855 | 1,830 - 1,881 | 0,00 |
| **Situação conjugal** | | | | | | |
| Sim | 1,000 | | | 1,000 | | |
| Não | 0,760 | 0,756 - 0,764 | 0,00 | 0,669 | 0,664 - 0,673 | 0,00 |
| **Raça** | | | | | | |
| Branca | 1,000 | | | 1,000 | | |
| Não Branca | 0,622 | 0,618 - 0,625 | 0,00 | 0,933 | 0,926 - 0,939 | 0,00 |
| **Escolaridade** | | | | | | |
| Ensino Básico | 1,000 | | | 1,000 | | |
| Ensino Superior | 0,967 | 0,958 - 0,975 | 0,00 | 1,172 | 1,160 - 1,184 | 0,00 |
| **Plano de saúde** | | | | | | |
| Sim | 1,000 | | | 1,000 | | |
| Não | 0,508 | 0,505 - 0,512 | 0,00 | 0,454 | 0,450 - 0,459 | 0,00 |

Fonte: Elaboração própria a partir dos microdados da PNAUM 2013/2014.

[a] Critério de Classificação Econômica Brasil (2013).

Legenda: AD – antidepressivos; NS/NR – não sabe/ não respondeu; IC – intervalo de confiança; OR – *odds ratio*

Na análise univariada, a variável mais fortemente associada ao uso de medicamentos AD foi a idade (OR = 8,128; IC 95% 8,010 – 8,247). Constatou-se também uma associação positiva entre o sexo feminino e o uso de AD, tendo as mulheres cerca de 70% de chances a mais de estarem em uso de medicamentos (OR = 1,691; IC 95% 1,681 – 1,701).

Em relação às regiões, encontrou-se uma probabilidade 4 (quatro) vezes maior de um indivíduo do Sul utilizar AD quando comparado a um indivíduo do Norte do Brasil (OR = 4,267; IC 95% 4,203 – 4,331). Quanto à classificação econômica, indivíduos da classe D/E apresentaram 26,3% menos chances (OR = 0,737; IC 95% 0,728 – 0,745) de estar em uso de medicamentos AD quando comparados àqueles pertencentes à classe B2.

Indivíduos que não estavam em uma situação conjugal apresentaram 24% menos chances (OR = 0,760; IC 95% 0,756 – 0,764) de estarem em uso de AD quando comparados àqueles que estavam em uma situação conjugal. Em relação à raça, a população não branca apresentou uma probabilidade 37,8% menor (OR = 0,622; IC 95% 0,618 – 0,625) de estar em uso de AD quando comparada à população branca.

Ao se comparar as probabilidades de uso de medicamentos antidepressivos (OR = 0,967; IC 95% 0,958–0,975), a população com ensino básico apresentou razão de chances muito próximas em relação à população com ensino superior. A parcela da população analisada que não possuía plano de saúde apresentou 49,2% (OR = 0,508; IC 95% 0,505 – 0,512) menos chances de estar em uso de AD.

Todas as variáveis analisadas relacionadas às características demográficas e socioeconômicas apresentaram *valor p* menor do que 0,20, sendo, portanto, incluídas na análise multivariada. Constatou-se uma maior associação positiva em relação ao uso de antidepressivos no sexo feminino (OR = 1,702; IC 95% 1,708 – 1,732), nos indivíduos com 60 anos ou mais (OR = 13,235; IC 95% 13,005 – 13,470) e naqueles da Região Sul (OR = 2,382; IC 95% 2,339 – 2,426).

O modelo de regressão logística foi estatisticamente significante, com $\chi^2(8) = 71.214, 744$, $p < 0,05$. O modelo explicou 10,7% (Nagelkerke $R^2$) da variação no uso de medicamentos antidepressivos e classificou corretamente 90,8% dos casos.

Na Tabela 6 estão descritas as razões de chances (OR) relacionadas à percepção da saúde dos indivíduos portadores de depressão em uso de AD.

Tabela 6 – Análises univariada (OR *não ajustada*) e multivariada (OR *ajustada*) dos fatores de percepção da saúde associados ao uso de AD. PNAUM 2013/2014.

| Variável e Categorias | Razão de chances (OR) não ajustada | IC 95% | P | Razão de chances (OR) ajustada | IC 95% | P |
|---|---|---|---|---|---|---|
| **Limitação causada pela depressão** | | | | | | |
| Não limita | 1,00 | | | 1,00 | | |
| Um pouco | 1,196 | 1,188 - 1,204 | 0,00 | 1,324 | 1,315 - 1,333 | 0,00 |
| Moderadamente | 0,836 | 0,830 - 0,842 | 0,00 | 0,930 | 0,923 - 0,937 | 0,00 |
| Intensamente | 2,508 | 2,477 - 2,539 | 0,00 | 2,712 | 2,678 - 2,746 | 0,00 |
| Muito intensamente | 1,927 | 1,989 - 1,957 | 0,00 | 1,994 | 1,963 - 2,026 | 0,00 |
| **Percepção da autoavaliação da saúde** | | | | | | |
| Muito boa | 1,00 | | | 1,00 | | |
| Boa | 0,155 | 0,151 - 0,159 | 0,00 | 0,146 | 0,142 - 0,150 | 0,00 |
| Regular | 0,154 | 0,150 - 0,158 | 0,00 | 0,133 | 0,130 - 0,137 | 0,00 |
| Ruim | 0,194 | 0,189 - 0,199 | 0,00 | 0,141 | 0,137 - 0,145 | 0,00 |
| Muito ruim | 0,380 | 0,368 - 0,392 | 0,00 | 0,278 | 0,269 - 0,287 | 0,00 |

Fonte: Elaboração própria a partir dos microdados da PNAUM 2014.
Legenda: AD – antidepressivos; NS/NR – não sabe/ não respondeu; IC – intervalo de confiança; OR – *odds ratio*

As características relacionadas à saúde também se apresentaram significativamente associadas ao uso de AD (p<0,05), sendo a limitação causada pela depressão positivamente associada ao uso de medicamentos.

### 4.2.2 Análise relacionada ao acesso a medicamentos AD por meio de farmácias públicas – SUS

A Tabela 7 apresenta os resultados das análises univariada e multivariada expressos em termos de razões de chances (OR) associadas ao acesso a medicamentos AD em relação às características demográficas e socioeconômicas.[4]

Tabela 7 – Análises univariada (OR não ajustada) e multivariada (OR ajustada) dos fatores demográficos e socioeconômicos associados ao acesso a AD por meio de farmácias públicas - SUS. PNAUM 2013/2014.

| Variável e Categorias | Razão de chances (OR) não ajustada | IC 95% | P | Razão de chances (OR) ajustada | IC 95% | P |
|---|---|---|---|---|---|---|
| **Sexo** | | | | | | |
| Masculino | 1,000 | | | 1,000 | | |
| Feminino | 1,182 | 1,177 - 1,187 | 0,00 | 1,264 | 1,257 - 1,270 | 0,00 |
| **Faixa Etária** | | | | | | |
| 15 a 19 anos | 1,000 | | | 1,000 | | |
| 20 a 59 anos | 2,103 | 2,078 - 2,128 | 0,00 | 3,897 | 3,835 - 3,959 | 0,00 |
| 60 anos ou mais | 1,853 | 1,830 - 1,875 | 0,00 | 3,524 | 3,468 - 3,582 | 0,00 |
| **Região do Brasil** | | | | | | |
| Norte | 1,000 | | | 1,000 | | |
| Nordeste | 0,970 | 0,959 - 0,982 | 0,00 | 0,951 | 0,938 - 0,964 | 0,00 |
| Sudeste | 1,573 | 1,555 - 1,591 | 0,00 | 2,893 | 2,855 - 2,931 | 0,00 |
| Sul | 1,735 | 1,714 - 1,755 | 0,00 | 2,910 | 2,870 - 2,950 | 0,00 |
| Centro-Oeste | 0,910 | 0,898 - 0,922 | 0,00 | 1,384 | 1,364 - 1,405 | 0,00 |
| **CCEB[a]** | | | | | | |
| Classe B2 | 1,000 | | | 1,000 | | |
| Classe C1 | 2,363 | 2,345 - 2,382 | 0,00 | 1,958 | 1,941 - 1,976 | 0,00 |
| Classe C2 | 3,825 | 3,797 - 3,854 | 0,00 | 3,063 | 3,037 - 3,089 | 0,00 |
| Classe D/E | 4,218 | 4,186 - 4,249 | 0,00 | 3,241 | 3,212 - 3,270 | 0,00 |
| **Situação conjugal** | | | | | | |
| Sim | 1,000 | | | 1,000 | | |
| Não | 1,040 | 1,036 - 1,043 | 0,00 | 0,946 | 0,942 - 0,950 | 0,00 |
| **Raça** | | | | | | |
| Branca | 1,000 | | | 1,000 | | |
| Não Branca | 1,208 | 1,204 - 1,212 | 0,00 | 1,394 | 1,387 - 1,400 | 0,00 |
| **Escolaridade** | | | | | | |
| Ensino Básico | 1,000 | | | 1,000 | | |
| Ensino Superior | 1,101 | 1,095 - 1,107 | 0,00 | 1,684 | 1,673 - 1,694 | 0,00 |

---

[4] Ressalta-se que, novamente, as categorias "classe A" e "classe B1" da variável *CCEB* foram excluídas, considerando que não houve casos registrados para a primeira categoria (0%) e a segunda representou apenas 0,06% da amostra. Desta forma, para efeito de melhor comparabilidade, optou-se pela exclusão das referidas categorias da análise, tomando a "classe B2" como a referência.

| | | | | | | |
|---|---|---|---|---|---|---|
| **Plano de saúde** | | | | | | |
| Sim | 1,000 | | | 1,000 | | |
| Não | 2,969 | 2,958 - 2,980 | 0,00 | 3,895 | 3,876 - 3,913 | 0,00 |

Fonte: Elaboração própria a partir dos microdados da PNAUM 2013/2014.

Critério de Classificação Econômica Brasil (2013).

Legenda: AD – antidepressivos; NS/NR – não sabe/ não respondeu; IC – intervalo de confiança; OR – *odds ratio*

Da mesma forma, a categoria "não sabe" da variável *plano de saúde* também foi excluída, considerando que não houve casos registrados (0%). Portanto, optou-se por excluí-la visando a comparação de interesse apenas entre aqueles indivíduos que possuíam e os que não possuíam convênio.

Na análise univariada, o fator mais fortemente associado positivamente ao acesso a medicamentos AD foi a classificação econômica (OR = 4,218; IC 95% 4,186 – 4,249), tendo um indivíduo da classe D/E 4 (quatro) vezes mais chances de ter acesso por meio do SUS ao AD do que àquele pertencente à classe B2.

A variável *plano de saúde* também apresentou forte associação positiva, indicando que o fato de não possuir o convênio eleva quase 3 (três) vezes as chances de acesso ao AD por meio do SUS (OR = 2,969; IC 95% 2,958 – 2,980).

Indivíduos entre 20 e 59 anos e aqueles acima de 60 anos apresentaram quase duas vezes mais chances de ter acesso ao AD por meio da farmácia pública quando comparados aos mais jovens, que possuem idade entre 15 e 19 anos. Sexo, situação conjugal, raça e escolaridade demonstraram fracas associações relacionadas ao acesso a medicamentos antidepressivos.

Todas as variáveis analisadas apresentaram nível de significância maior que 0,20 (p > 0,20) e, portanto, foram incluídas na análise multivariada. Como resultado dessa análise, os fatores que apresentaram associação positiva mais forte foram a idade (OR = 3,897; IC 95% 3,835 – 3,959), o plano de saúde (OR = 3,895; IC 95% 3,876 – 3,913) e a classificação econômica (OR = 3,24; IC 95% 3,212 – 3,270).

O modelo de regressão logística foi estatisticamente significante, $\chi^2(14) = 681420,765$, $p < 0,05$, explicando 18,4% (Nagelkerke $R^2$) da variação no acesso a medicamentos antidepressivos e classificou corretamente 69,7% dos casos.

**6. Comentários Finais**

De maneira geral, os resultados evidenciam uma alta prevalência de uso de medicamentos antidepressivos nos indivíduos portadores de depressão no Brasil. O perfil da população brasileira em uso de medicamentos antidepressivos apresentou prevalência de indivíduos do sexo feminino, entre 20 e 59 anos, brancos, da Região Sudeste, da classe econômica D/E, com nível médio de escolaridade, em uma situação conjugal e sem cobertura de plano de saúde. Além disso, a população em estudo relatou não apresentar limitações derivadas da depressão e autoavaliou sua saúde como regular.

Ao se traçar o perfil de uso de medicamentos antidepressivos verifica-se que este é idêntico ao perfil de acesso a AD por meio de farmácias públicas do SUS. Nota-se também que o fator idade demonstrou associação positiva tanto para o uso quanto para o acesso a antidepressivos. Os outros fatores demográficos e socioeconômicos que mais se associaram ao uso dos medicamentos antidepressivos foram o sexo e a região do Brasil. Quanto ao acesso, posse de plano de saúde e a classificação econômica demonstraram uma forte associação.

**REFERÊNCIAS**